\documentclass[12pt]{article}
\usepackage{graphicx}
\newcommand\pubnumber{\ }
\newcommand\pubdate{\today}

\def\institute{CERN, EP department, CH-1211 Geneva 23, Switzerland}
\def\support{\footnote{On behalf of the ATLAS and CMS Collaborations}}

\def\Title#1{\begin{center} {\Large #1 } \end{center}}
\def\Author#1{\begin{center}{ \sc #1} \end{center}}
\def\Address#1{\begin{center}{ \it #1} \end{center}}

\newcommand\pubblock{\rightline{\begin{tabular}{l} \pubnumber\\
         \pubdate  \end{tabular}}}
\newenvironment{Abstract}{\begin{quotation}  }{\end{quotation}}
\newenvironment{Presented}{\begin{quotation} \begin{center} 
             PRESENTED AT\end{center}\bigskip 
      \begin{center}\begin{large}}{\end{large}\end{center} \end{quotation}}

\def\beq{\begin{equation}}
\def\eeq#1{\label{#1}\end{equation}}
\def\eeqn{\end{equation}}

\def\beqa{\begin{eqnarray}}
\def\eeqa#1{\label{#1}\end{eqnarray}}
\def\eeqan{\end{eqnarray}}

\let\bar=\overbar

\def\Dslash{\not{\hbox{\kern-4pt $D$}}}
\def\dslash{\not{\hbox{\kern-2pt $\del$}}}

\def\msb{{\bar{\ssstyle M \kern -1pt S}}}

\newcommand{\ttbar}{\mbox{$t\bar{t}$}}
\newcommand{\qqbar}{\mbox{$q\bar{q}$}}
\newcommand{\bbbar}{\mbox{$b\bar{b}$}}
\newcommand{\xtt}{\mbox{$\sigma(\ttbar)$}}
\newcommand{\mtop}{\mbox{$m_{t}$}}
\newcommand{\mtpole}{\mbox{$m_{t}^{\rm pole}$}}

\newcommand{\sxyt}{$\sqrt{s}=13$\,TeV}

\newcommand{\aconfref}[3]{ATLAS Collaboration, ATLAS-CONF-#2}
\newcommand{\arefpaper}[3]{ATLAS Collaboration, #2, arXiv:#3}

\newcommand{\aplb}[3]{\arefpaper{#1}{Phys.\ Lett.\ B#2}{#3}}
\newcommand{\ajhep}[3]{\arefpaper{#1}{JHEP #2}{#3}}
\newcommand{\aepjc}[3]{\arefpaper{#1}{Eur.\ Phys.\ J.\ C#2}{#3}}

\newcommand{\jhep}[4]{#1, JHEP #3, arXiv:#4}
\newcommand{\epjc}[4]{#1, Eur.\ Phys.\ J.\ C#3, arXiv:#4}

\newcommand{\cpc}[4]{#1, Comp.\ Phys.\ Comm.\ #3, arXiv:#4}

\newcommand{\prab}[3]{#1, Phys.\ Rev.\ Accel.\ Beams #3}

\newcommand{\arxiv}[3]{#1, arXiv:#3}

\begin{document}
\begin{titlepage}
\pubblock

\vfill
\Title{Inclusive \ttbar\ cross-section measurements at LHC}
\vfill
\Author{ Richard Hawkings\support}
\Address{\institute}
\vfill
\begin{Abstract}
A review of ATLAS and CMS measurements of the inclusive \ttbar\
production cross-section in $pp$ collisions at $\sqrt{s}=7$--13\,TeV
is presented, focusing on the most precise results in the dilepton
and lepton+jets final states. The measurements are in good agreement with 
state-of-the-art QCD predictions, and have been used to determine
the top quark pole mass and provide constraints on proton parton distribution
functions.
\end{Abstract}
\vfill
\begin{Presented}
$10^{th}$ International Workshop on Top Quark Physics\\
Braga, Portugal,  September 18--22, 2017
\end{Presented}
\vfill
\end{titlepage}
\def\thefootnote{\fnsymbol{footnote}}
\setcounter{footnote}{0}

\section{Introduction}

The inclusive top-pair (\ttbar) production cross-section \xtt\ is an important
quantity for characterising $pp$ collisions in the TeV energy regime. At typical
LHC energies ($\sqrt{s}=$\,7--13\,TeV), \ttbar\ production is dominated
by gluon fusion. It has been calculated to next-to-next-to-leading-order
(NNLO) precision with resummation of next-to-next-to-leading-logarithmic
(NNLL) soft gluon terms \cite{toppp}. The uncertainties on the predictions
for \xtt\ at a fixed top quark mass of $\mtop=172.5$\,GeV 
are around 5\,\%,  dominated by parton distribution 
function (PDF) and QCD scale uncertainties. The predicted cross-section also
decreases by 3\,\% for a 1\,GeV increase in \mtop.

\begin{figure}[tp]
\vspace{-5mm}
\hspace{20mm}\includegraphics[width=100mm]{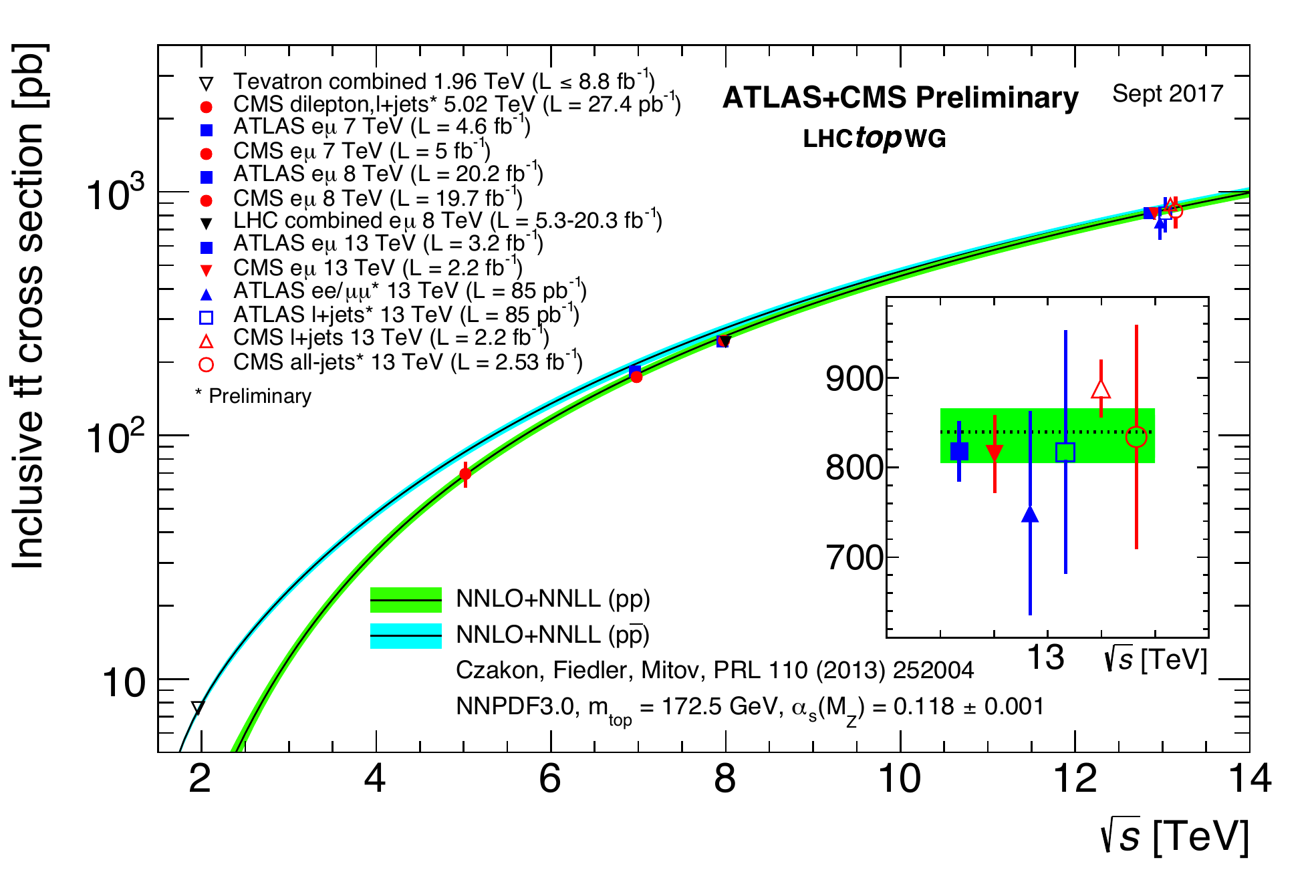}
\vspace{-7mm}
\caption{\label{f:xsecvsecm}Selected LHC and Tevatron \ttbar\ cross-section
measurements as a function of $\sqrt{s}$ compared to
NNLO+NNLL QCD predictions \cite{topplots}.}
\vspace{-3mm}
\end{figure}

The most precise measurements of \xtt\ from the ATLAS \cite{atlasdet}
and CMS \cite{cmsdet} experiments make use of the dilepton
($\ttbar\rightarrow\ell^+\ell^-\nu\bar{\nu}\bbbar$) and lepton+jets
($\ttbar\rightarrow\ell^{\pm}\nu\qqbar\bbbar$) channels, where both $W$ bosons 
or just one decay to an electron or muon (denoted $\ell$) and a neutrino.
Less precise
measurements in the all-hadronic channel and in final states involving
hadronically-decaying $\tau$-leptons have also been reported. The measurements
are in good agreement with the predictions at all energies, as shown 
in Figure~\ref{f:xsecvsecm}; when also including Tevatron results in $p\bar{p}$
collisions at $\sqrt{s}=1.96$\,TeV, the comparison spans two orders of magnitude
in \xtt.

\section{Measurement techniques}

The cleanest measurements have been made in the dilepton channel with
one electron and one muon, thereby suppressing the background from 
$Z\rightarrow\ell\ell$+jets with two same-flavour leptons. 
The remaining background
is dominated by $Wt\rightarrow e\mu$+($b$)jets events at the level of 3--10\,\%
depending on the selection.
The CMS 
$\sqrt{s}=13$\,TeV analysis \cite{mclly} selected events with at least two jets
(at least one being $b$-tagged), and determined \xtt\ from the yield of 
events after 
background subtraction. The resulting systematic uncertainties due to
modelling of light jet production in \ttbar\ events, jet energy scale and
$b$-tagging efficiency were reduced in the ATLAS analyses at 7, 8 and 13\,TeV
\cite{mallv,mally} by focusing on $b$-tagged jets only, 
and fitting simultaneously \xtt\ and the
probability to reconstruct and tag a $b$-jet from the top decay, 
from the rates of events with one and two $b$-tagged jets.
The 7--8\,TeV CMS analysis \cite{mcllv} went one step further, also
using the zero $b$-tag events, and fitting the untagged jet multiplicity
and jet transverse momentum distributions to constrain systematic uncertainties.

\begin{figure}[tp]
\vspace{-5mm}
\includegraphics[width=95mm]{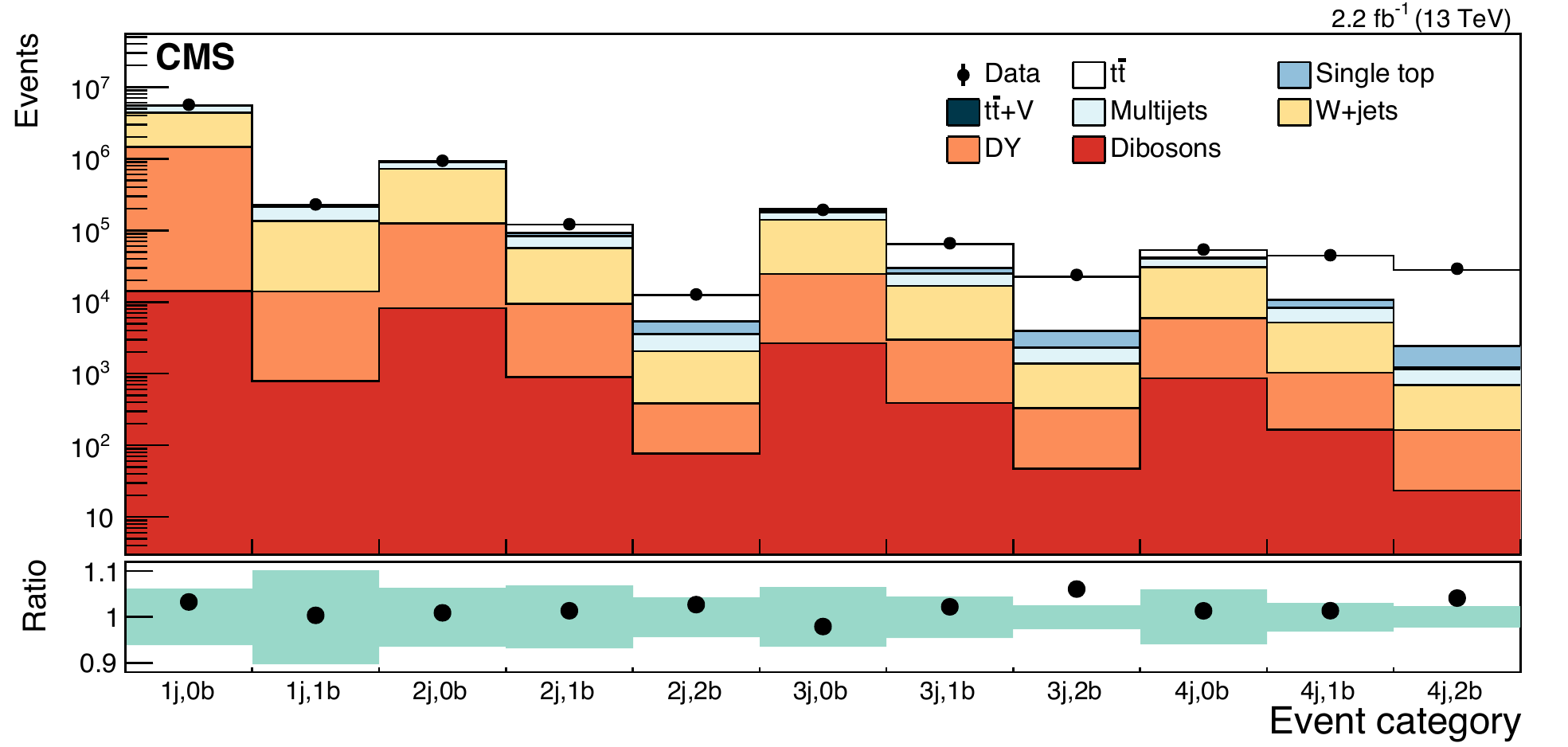}
\includegraphics[width=48mm]{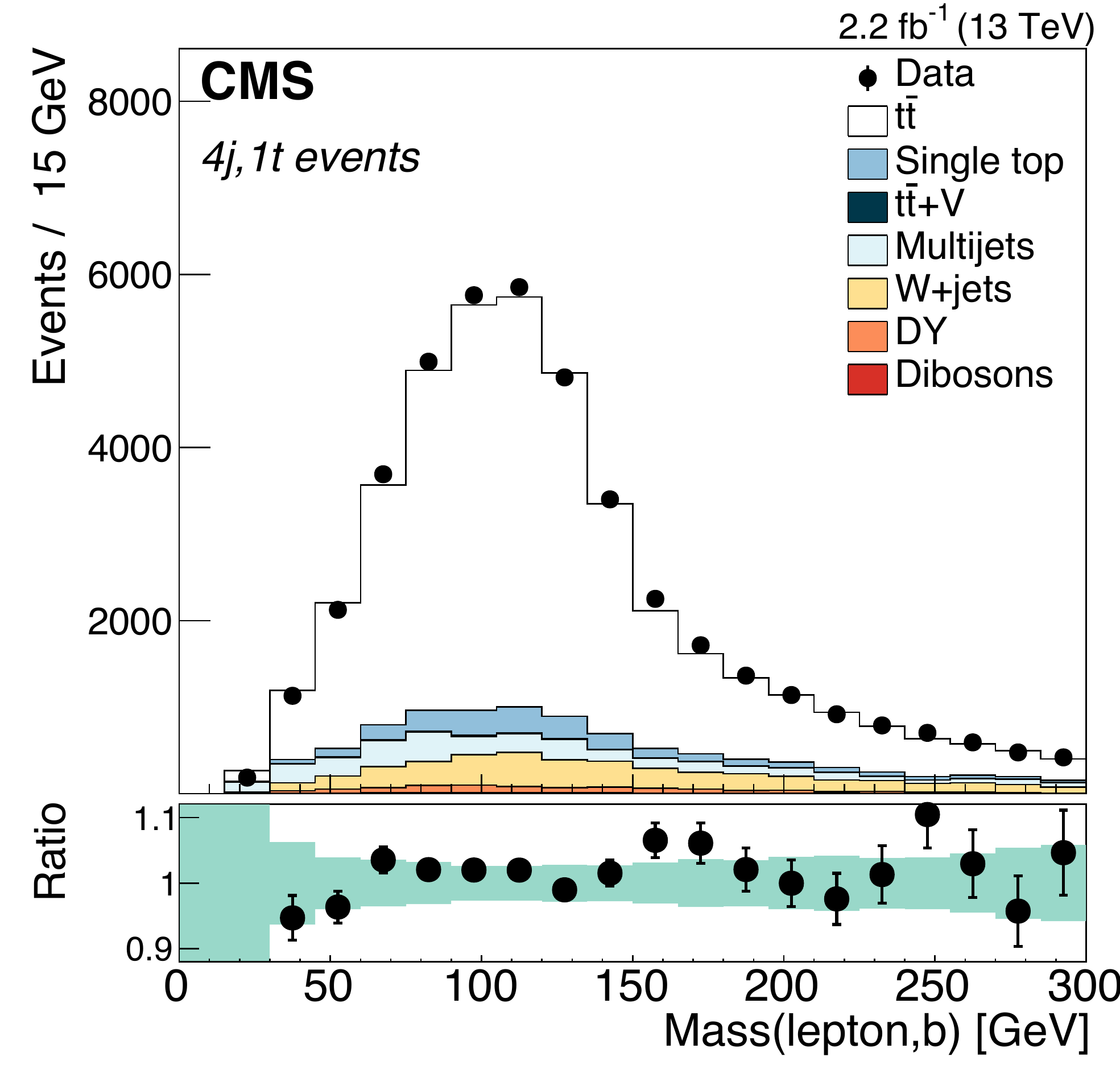}
\vspace{-3mm}
\caption{\label{f:cmsljet}Event yields in various jet/$b$-tagged jet 
multiplicity categories (left) and invariant mass of the lepton and 
$b$-tagged jet (right) for four jet/one $b$-tag events in the 
CMS lepton+jets analysis at $\sqrt{s}=13$\,TeV \cite{mcljy}.}
\vspace{-3mm}
\end{figure}

The lepton+jets channel offers larger statistics but brings significant
additional backgrounds from $t$-channel single top,
$W$+jet and QCD multijet production, that are typically estimated using
data-driven techniques. The latest ATLAS $\sqrt{s}=8$\,TeV analysis
\cite{maljv} used the reconstructed two-jet invariant mass peak 
from the hadronic
$W\rightarrow\qqbar$ decay and the rates of events with different jet and 
$b$-tagged jet multiplicities to reduce the jet energy scale and $b$-tagging
uncertainties. It also made use of data $Z\rightarrow\ell\ell$+jets events
with one lepton transformed into a neutrino, in order to better model
the $W$+jets background. The $\sqrt{s}=13$\,TeV CMS analysis \cite{mcljy}
used a fit to all events with a lepton, significant missing transverse momentum
and at least one jet, and categorised them according to jet and $b$-tagged
jet multiplicity. As shown in Figure~\ref{f:cmsljet}, the low multiplicity
bins are dominated by backgrounds, allowing their normalisation to be 
determined from a fit, which also made use of a discriminating distribution
({\em e.g.} the invariant mass of the lepton and $b$-tagged jet) in each
multiplicity category. The relative contributions of each event source
in each category and bin were described by templates, with the effects of
systematic uncertainties on these templates from detector and
physics modelling effects being accounted for via
auxiliary fit (`nuisance') parameters. The large number of fit categories
with varying signal contributions allowed these parameters to be
strongly constrained by the data,
under the assumption that the model correctly captures all the sources
of systematic uncertainty and their correlations across different bins and
distributions.

\section{Comparison of uncertainties}

The fractional uncertainties of the most precise dilepton and lepton+jets
measurements at $\sqrt{s}=8$ and 13\,TeV are compared in Table~\ref{t:uncsum},
classifying the uncertainties reported in the original publications
into various categories. The LHC beam energy-related uncertainties
quoted in Ref. \cite{mallv,mally} have been neglected, given that the 
beam energy has now been determined to a precision of 0.1\,\% \cite{lhcbeam},
corresponding to an effect of 0.2--0.3\,\% on \xtt.

\begin{table}
\centering

\begin{tabular}{rrl|ccccc|c}\hline
$\sqrt{s}$  & $\int L$ & Analysis & Stat. & \ttbar\ mod. & Det. & Bkg. & Lumi. & Total \\
(TeV) & (fb$^{-1}$) & & (\%) & (\%) & (\%) & (\%) & (\%) & (\%) \\
\hline
13 & 3.2  & ATLAS $\ell\ell$ \cite{mally} & 0.9 & 3.0 & 1.1 & 0.9 & 2.3 & 4.1 \\
13 & 2.2  & CMS $\ell\ell$ \cite{mclly}   & 1.0 & 2.4 & 3.6 & 1.5 & 2.3 & 5.3 \\
13 & 2.2  & CMS $\ell$+jets \cite{mcljy}  & 0.2 & 1.7 & 2.3 & 1.9 & 2.3 & 3.8 \\
 8 & 20.2 & ATLAS $\ell\ell$ \cite{mallv} & 0.7 & 1.7 & 1.2 & 0.9 & 2.1 & 3.2 \\
 8 & 19.7 & CMS $\ell\ell$ \cite{mcllv}   & 0.6 & 1.3 & 2.2 & 1.5 & 2.6 & 3.7 \\
 8 & 20.2 & ATLAS $\ell$+jets \cite{maljv}& 0.3 & 4.1 & 2.3 & 1.3 & 1.9 & 5.7 \\
 8 & 19.6 & CMS $\ell$+jets \cite{mcljv}  & 1.6 & 5.4 & 2.5 & 0.2 & 2.6 & 6.5 \\
\hline
\end{tabular}
\caption{\label{t:uncsum}Comparison of fractional uncertainties
due to data statistics, \ttbar\ modelling, detector effects, 
background modelling and
integrated luminosity, and the total uncertainty, for the most precise
\xtt\ measurements at $\sqrt{s}=8$ and 13\,TeV.}
\end{table}

At $\sqrt{s}=8$\,TeV, dilepton measurements have the smallest uncertainties,
with lepton+jets results suffering in particular from larger \ttbar\ modelling
uncertainties. At 13\,TeV, the early dilepton results have significantly
larger uncertainties than at 7--8\,TeV, which should be
reducible once the results of Monte Carlo tuning studies based on 13\,TeV
data are fully incorporated, and the lepton efficiencies, energy scales
and resolutions are fully characterised using the copious 
$Z\rightarrow\ell\ell$ data. The most precise 13\,TeV result to date
comes from CMS in the lepton+jets channel \cite{mcljy}, and benefits from 
the strong reduction in \ttbar\ modelling uncertainties due to the in-situ 
constraints from the  fit, to a much greater degree than in previous
7--8\,TeV results.

\section{Interpretation and outlook}

\begin{figure}
\includegraphics[width=72mm]{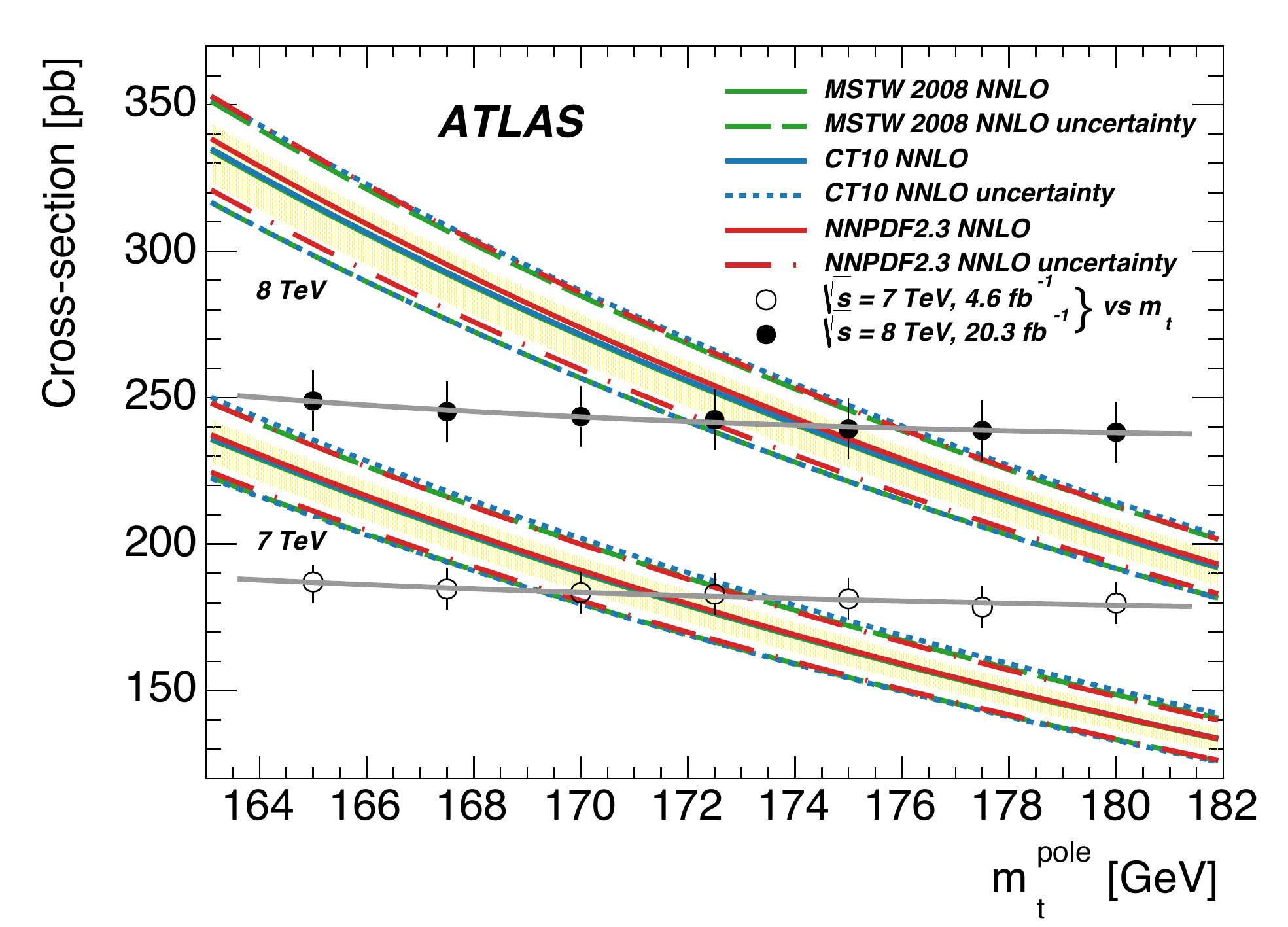}
\includegraphics[width=80mm]{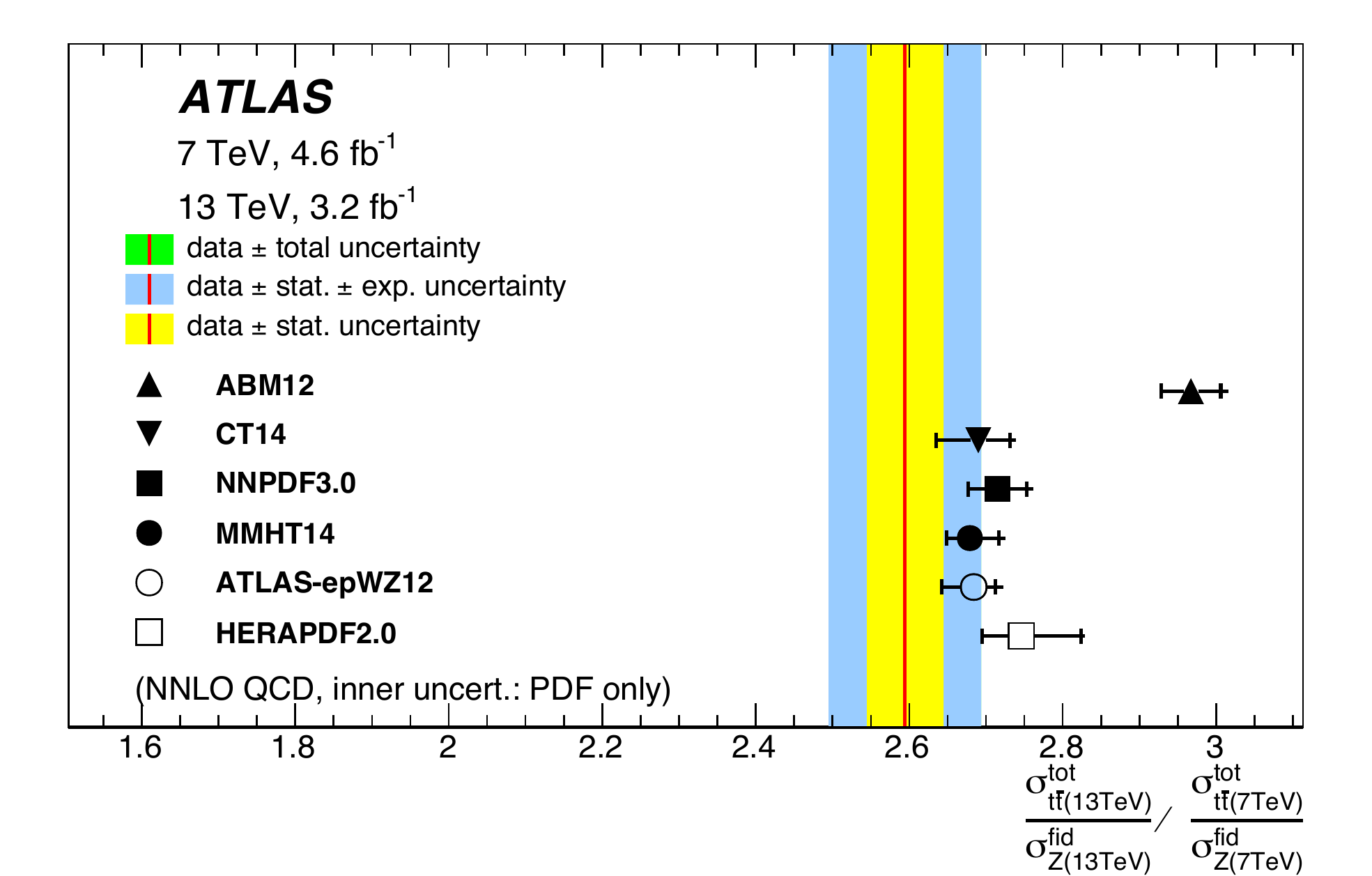}
\vspace{-8mm}
\caption{\label{f:mtopratio}Comparison of ATLAS $\sqrt{s}=7$ and 8\,TeV
measurements of \xtt\ {\em vs.} assumed \mtop\ with NNLO+NNLL predictions for 
various PDF sets \cite{mallv} (left); measured double ratios of \ttbar\ and 
$Z\rightarrow\ell\ell$ cross-sections at $\sqrt{s}=13$ and 7\,TeV
compared to predictions from various PDF sets \cite{azratio} (right).}
\vspace{-2mm}
\end{figure}

The theoretical predictions for \xtt\ are dependent on the top quark pole
mass \mtpole, a well-defined mass definition corresponding to that of a
free particle, without some of the ambiguities inherent in reconstructing
\mtop\ from its decay products in a hadron collider environment. Some of the
precise \xtt\ measurements have therefore been interpreted as measurements
of \mtpole, as summarised in Table~\ref{t:mtopxsec}. Here, it is important
to take into account the residual dependence of the measured \xtt\ on \mtop\
through \ttbar\ acceptance and single top-quark production background
(see Figure~\ref{f:mtopratio} (left)).
This technique can reach a precision of about $\pm 2$\,GeV, driven mainly
by PDF-related uncertainties on the predictions. 
It is important to
use PDFs based on datasets which do not include \ttbar\ production data
to avoid potential circularity (the determination based on NNPDF3.0 in 
Table~\ref{t:mtopxsec} has the smallest uncertainties, but NNPDF3.0 includes
some constraints based on earlier LHC \xtt\ measurements). The CMS
measurements also treat the QCD scale uncertainties on the predictions
with a uniform prior distribution, whereas ATLAS uses a more conservative 
Gaussian distribution.

\begin{table}
\centering
ß
\begin{tabular}{rl|l|l}\hline
$\sqrt{s}$ (TeV) & Analysis & PDF set & \mtpole (GeV) \\
\hline
7+8 & ATLAS $\ell\ell$ \cite{mallv} & PDF4LHC Run-1 & $172.9^{+2.5}_{-2.6}$ \\
7+8 & CMS $\ell\ell$ \cite{mcllv} & NNPDF3.0 & $173.8^{+1.7}_{-1.8}$ \\
7+8 & CMS $\ell\ell$ \cite{mcllv} & MMHT2014 & $174.1^{+1.8}_{-2.0}$ \\
7+8 & CMS $\ell\ell$ \cite{mcllv} & CT14 & $174.3^{+2.1}_{-2.2}$ \\
13 & CMS $\ell$+jets \cite{mcljy}  & CT14 & $170.6\pm 2.7$ \\
\hline
\end{tabular}
\caption{\label{t:mtopxsec}Measurements of the top quark pole mass \mtpole\
from the inclusive \ttbar\ production cross-section \xtt\ 
for various datasets and PDF choices.}
\end{table}

Predictions for ratios of \ttbar\ cross-sections $R(A/B)$ at different 
$\sqrt{s}$ values $A$ and $B$
benefit from significant uncertainty cancellations, {\em e.g.} 
$R(8/7)$ is predicted to be $1.430\pm 0.013$. Some systematics also cancel
in the experimentally-measured ratios, especially for measurements
performed with the same technique at different $\sqrt{s}$ values. The most
precise measurement is currently $R(8/7)=1.328\pm 0.047$ \cite{mallv},
$2.1\sigma$ below the prediction and with an uncertainty dominated by 
the uncertainties on the integrated luminosities, which are only weakly 
correlated between the 7 and 8\,TeV samples.  Other $R(8/7)$ results 
\cite{mcllv,mcljv} and
first ratios including $\sqrt{s}=13$\,TeV data \cite{azratio} are in 
good agreement with the predictions. Smaller uncertainties can be achieved
by forming double-ratios of \ttbar\ and $Z$-boson cross-sections at 
different energies, where the normalisation via the $Z$ exchanges the luminosity
uncertainty for a dependence on the quark/anti-quark PDF uncertainties
which drive the $Z$ cross-section predictions. First measurements of
double ratios involving $\sqrt{s}=13$\,TeV data 
(Figure~\ref{f:mtopratio} (right)) show some sensitivity to PDFs, 
but would benefit from improved 13\,TeV \ttbar\ results with precision
matching those at 7--8\,TeV.

In conclusion, the measurements of \ttbar\ inclusive cross-sections
at LHC are now quite mature, with several precise (3--4\,\%) results at 
$\sqrt{s}=7$--8\,TeV from both ATLAS and CMS. Further improvements
can be expected at 13\,TeV with the use of larger datasets and refined
analyses, and the 
first limited-precision measurement at 5\,TeV \cite{mcljq} with
27\,pb$^{-1}$ should be followed-up using the larger dataset from 2017. The
results to date are generally in good agreement with the predictions
from NNLO+NNLL QCD, and have been used to constrain PDFs and determine
the top quark pole mass.

\end{document}